\documentclass[letterpaper]{JHEP3}

\usepackage[dvips]{graphicx}

\newcommand{\arr[1]}{ -\!\!\! \longrightarrow^{\hskip-.27in #1}\ }

\title{Low-scale gaugino mediation, lots of leptons at the LHC}

\author{Andrea De Simone\\
        Center for Theoretical Physics,
        Massachusetts Institute of Technology, Cambridge, MA 02139\\
        E-mail: \email{andreads@mit.edu}}
\author{JiJi Fan\\
        Department of Physics, Sloane Laboratory, Yale University, New Haven, CT 06520\\
        E-mail: \email{jiji.fan@yale.edu}}
\author{Martin Schmaltz\\
        Physics Department, Boston University, 590 Commonwealth Ave, Boston,
        MA 02215\\
        E-mail: \email{schmaltz@bu.edu}}
\author{Witold Skiba\\
        Department of Physics, Sloane Laboratory, Yale University, New Haven, CT 06520\\
        E-mail: \email{witold.skiba@yale.edu}}

\abstract{Low-scale gaugino mediation predicts that gauginos are
significantly heavier than scalar superpartners. In order of
increasing mass the lightest superpartners are the gravitino,
right-handed sleptons and left-handed sleptons (no light
neutralino!). This implies that squark decay chains pass through
one or more sleptons and typical final states from squark and
gluino production at the LHC include multiple leptons. In
addition, left-handed staus have large branching fractions into
right-handed staus and the Higgs. As an example, we compute the
spectrum of low-scale deconstructed gaugino mediation. In this
model gauginos acquire masses at tree level at 5 TeV while scalar
masses are generated radiatively from the gaugino masses.}

\newcommand{\eqn}[1]{\label{eq:#1}}
\newcommand{\refeq}[1]{(\ref{eq:#1})}

\newcommand{\Eq}{Eq.~\refeq}

\newcommand{\beq}{\begin{eqnarray}}
\newcommand{\eeq}{\end{eqnarray}}

\begin{document}

\section{Introduction}

The LHC will soon start collecting data and much effort will be
devoted to searching for supersymmetry. The search strategies depend on the
superpartner spectrum. Given the multitude of soft supersymmetry-breaking parameters
one turns to models for guidance on which spectra are reasonable possibilites.
In this paper, we compute the spectrum of low-scale gaugino mediation, a model which is obtained as a limit of gaugino mediation ~\cite{Kaplan:1999ac} and which predicts that scalar superpartners are much lighter than gauginos. As we will show, this leads to some interesting new collider signatures.

What is unusual about the spectrum of low-scale gaugino mediation
is that both left- and right-handed sleptons are lighter than all
other superpartners except the gravitino. As a consequence, all
superpartner decay chains pass through one or two sleptons,
leading to multiple leptons in final states. For example, a pair
of squarks produced at the LHC decays into 2 jets and 5 leptons
(not counting $\tau$ leptons or neutrinos) about 1\% of the time.
In addition, we expect two heavy charged tracks from the
right-handed sleptons which, for most of parameter space, pass
through the detector without decaying. This is quite different
from gaugino mediation at high scales \cite{Schmaltz:2000gy} where
typically the Bino is the Lightest Supersymmetric Particle (LSP).
A similar phenomenology occurs in gauge
mediation~\cite{Dine:1994vc} with light right-handed sleptons
except that low-scale gaugino mediation predicts even more leptons
in decay chains from the additional light left-handed sleptons.
Another interesting feature of our spectrum is that the
left-handed stau decays to the right-handed stau and a Higgs with
large branching fraction. Since these events are easily
distinguished from QCD backgrounds by the stable charged tracks
and leptons one would be able to discover the Higgs rather easily
in the $b\overline b$ channel.

Gaugino mediation arises when supersymmetry breaking generates gaugino masses without simultaneously generating significant scalar masses at the ``messenger scale''~\cite{Kaplan:1999ac}. The scalar masses are then obtained radiatively from the gaugino masses.  These radiative contributions can be divided into threshold effects at the messenger and the gaugino mass scales, and the logarithmic running between these two scales. In most other models, the logarithms are large so that scalar masses can catch up to gaugino masses and threshold effects are subdominant. The distinctive feature of low-scale gaugino mediation is that the messenger scale is near the superpartner masses scale $\sim$ TeV so that the logarithm is small. As a consequence, scalar masses remain significantly smaller than gaugino masses and one needs to include both high- and low-scale threshold effects to reliably compute the scalar masses. While the low-scale threshold effects only depend on the MSSM spectrum and can be computed without any further assumptions, the high-scale threshold effects depend on model-dependent messenger physics.

As an illustration, we construct an explicit model of low-scale gaugino
mediation based on ``deconstructed gaugino mediation'' \cite{Cheng:2001an,Csaki:2001em}.
In deconstructed gaugino mediation, each gauge group of the minimal supersymmetric standard model (MSSM) is the diagonal subgroup of two gauge groups $G_A$ and $G_B$ in the ultraviolet. The breaking to the diagonal is accomplished by a vacuum expectation value $\left<\Phi\right>\sim 5$ TeV of a bi-fundamental link field $\Phi$. If one further assumes that supersymmetry breaking only couples to $G_B$ whereas the MSSM matter fields are only charged under $G_A$, then negligibly small scalar masses result at high scales. Once the gauge groups $G_A$ and $G_B$ break to the diagonal, the MSSM gauginos obtain a tree-level mass. At the same threshold the MSSM scalars obtain a much smaller mass from loops involving the gauginos and link fields. Thus in this model the unspecified ``messenger physics'' of the previous paragraph is realized with heavy gauge bosons, gauginos and link fields.

In Section 3 we explicitly compute the leading contributions to
the gaugino and scalar masses in our model. While all MSSM matter
superpartners obtain positive masses squared we find that the
$H_u$ Higgs field has a negative mass-squared due to a
top-stop-gluino 2-loop diagram which is enhanced by color factors
and large coupling constants. Electroweak symmetry breaks
naturally because of this negative mass-squared. We also find that
the product group structure provides additional D-terms which
enhance the Higgs mass \cite{Batra}.

Using our calculation for the soft masses we determine the spectrum of superpartners and initiate a phenomenological study in Section 4. We find large branching fractions to leptons at the LHC, very mild limits on parameter space from precision electroweak measurements, and some contraints on the gravitino mass from cosmological considerations.
In Sec. 5, we point to outstanding questions and
discuss possible extensions of our work. Appendix A contains the details of our spectrum computations.

\section{A model of low-scale gaugino mediation}
\label{model}

Our model consists of the gauge group $SU(3)\times SU(2)\times U(1)$ with the
usual MSSM matter content and Higgs fields. For brevity of notation, we will often think of this gauge group as a subgroup of an $SU(5)$ which we call $SU(5)_{A}$. This allows us to specify the matter content of the model in terms of multiplets of $SU(5)_{A}$.

The model also contains an $SU(5)_{B}$ gauge group and adjoint chiral superfield $A$ of $SU(5)_B$. There is also a pair of bi-fundamental ``link fields" $\Phi$ and $\bar \Phi$. $\Phi$ transforms as a fundamental under $SU(5)_A$ and an antifundamental under $SU(5)_B$, while $\bar \Phi$ is an anti-fundamental under  $SU(5)_A$ and a fundamental under $SU(5)_B$.
Finally, we require a gauged $U(1)_B$ symmetry under which $\Phi$ and $\bar \Phi$ carry opposite charges, and a gauge singlet chiral superfield $S$. The field content and representations are summarized in the following table:
\beq
\begin{array}{c|c|c}
& SU(3)\!\times\! SU(2) \!\times\! U(1) \subset SU(5)_A & SU(5)_B \times U(1)_B \\
\hline
Q,U,D,L,E,H_u,H_d & ${ as in MSSM }$  & 1 \\
\Phi & \textbf{5} & \overline{\textbf{5}} \\
\bar \Phi & \overline{\textbf {5}} & \textbf{5} \\
A + S & 1 & \textbf{24} + 1 \\
\end{array}
\eeq

In addition to the MSSM superpotential, our model has the following superpotential
\beq
W= \lambda_A A \Phi \bar \Phi + \lambda_S S (\Phi \bar \Phi -  \Lambda^2)\, ,
\eqn{superpot}
\eeq
where $\lambda_A$ and $\lambda_S$ are coupling constants of order unity.%
\footnote{We have chosen the couplings $\lambda_A$ and $\lambda_S$ to be $SU(5)_A$ symmetric. The fact that not the full $SU(5)_A$ group is gauged will lead to splittings of the $\lambda's$ due to renormalization. However, the effect of these splittings on the superpartner mass spectrum in our model is only logarithmic and numerically not very significant. We will ignore these effects throughout.}
At the scale $\Lambda$, which we take to be $\sim 5$ TeV,%
\footnote{Requiring this scale to be close to the superpartner mass scale represents a tuning. A  spectrum similar to ours is obtained if the symmetry breaking VEVs of the link fields are induced by negative soft masses for these fields, thereby dynamically linking the scale of superpartner masses to the scale of the link field VEVs.}
the link fields obtain vacuum expectation values (VEVs)
\beq
\langle \Phi \rangle = \langle \bar \Phi \rangle \equiv f\, \bf{1} \, ,
\eeq
breaking the $SU(5)_A\times SU(5)_B$ symmetry to the diagonal subgroup.
Ignoring supersymmetry breaking,
the VEV is given by $f=\Lambda$. The super Higgs mechanism gives mass to
a linear combination of the A and B gauge multiplets, so that only the ``diagonal"
$SU(3)\times SU(2) \times U(1)$ gauge bosons remain light. The corresponding gauge couplings are related to the gauge couplings of the unbroken theory by
\beq
\frac{1}{g_i^2}=\frac{1}{g_{A_i}^2}+\frac{1}{g_B^2}\, ,
\eeq
where $i=1,2,3$ labels the groups $U(1), SU(2), SU(3)$, respectively.
It will also be convenient to define  the mixing angles $\theta_i$  by the
functions
\beq
s_i\equiv \sin\theta_i = \frac{g_{A_i}}{\sqrt{g_{A_i}^2+g_B^2}} \ , \quad \quad
c_i\equiv \cos\theta_i=\frac{g_B}{\sqrt{g_{A_i}^2+g_B^2}} \ .
\eeq
The gauge bosons with masses of order $f$ fill out an adjoint of $SU(5)_{\textrm{\scriptsize{diagonal}}}$.
Ignoring supersymmetry and electroweak symmetry breaking, their masses are given by
\beq
M^2_{i}&=&2 f^2 (g_B^2+g_{A_i}^2) \ , \nonumber\\
M^2_{X,Y}&=&2f^2 (g_B^2)\ .
\eeq
All chiral superfields $A, S, \Phi, \bar \Phi$ obtain masses. The linear combination $\Phi_{+}=(\Phi+\bar \Phi)/\sqrt{2}$ obtains a Dirac mass $ \lambda f$ with $A+S$ from the superpotential \Eq{superpot}, and the combination $\Phi_{-}=(\Phi-\bar\Phi)/\sqrt{2}$ is ``eaten" by the super Higgs mechanism.

The MSSM matter fields inherit couplings to the light MSSM gauge bosons from their couplings to the A gauge fields. They also have couplings to the heavy gauge multiplets which we will discuss further when we examine precision electroweak constraints in Section \ref{pheno}.

Let us now turn to discuss supersymmetry breaking.
We assume that supersymmetry breaking only couples to the $SU(5)_B$ gauge fields and any fields charged under $SU(5)_B$. There are several ways to achieve this scenario. For example, we could have gauge mediation with messengers charged under $SU(5)_B$ but not under $SU(5)_A$. Alternatively, one could have separation of A and B fields in extra dimensions with
supersymmetry breaking being localized near $SU(5)_B$. We will take a phenomenological approach and introduce independent arbitrary soft masses $m_B$ for the $SU(5)_B$ gauginos and a common soft mass squared $m^2_\Phi$ for the link fields.
Soft masses for the adjoint $A$ and the singlet $S$ will not be relevant for our analysis.
The MSSM matter fields and A gauginos also obtain small masses at two loops from diagrams involving both the A and B gauge interactions. However, these masses are negligible compared to the masses which we are about to compute from physics at and below the symmetry-breaking scale $f$.

After the link fields develop their vacuum expectation values, the gauginos of the A and B gauge groups mix and the supersymmetry-breaking gaugino mass $m_B$ contributes to both mass eigenstates. More precisely, there are three two-component adjoint fermions for each MSSM gauge generator (the two gauginos and the ``eaten" fermion from $\Phi_{-}$) which mix via a $3\times3$ mass matrix
\beq
\Omega_i=\pmatrix{0&0&g_{A_i}\sqrt{2} f \cr 0&m_B&-g_B\sqrt{2} f \cr g_{A_i} \sqrt{2}f& -g_B \sqrt{2}f &0}=
M_i \pmatrix{0&0&s_i \cr 0&{m_B}/{M_i}&-c_i \cr s_i& -c_i&0} \, .
\eqn{gauginomatrix}
\eeq
In the limit of small supersymmetry breaking ($m_B \ll M_i$) the light gaugino mass eigenstates are the superpartners of the MSSM gauge bosons with masses given by
\beq
m_i\simeq s_i^2 \, m_B= {g_i^2}\ \frac{m_B}{g_B^2}\, .
\eeq
Note this implies the usual gaugino mass ratios of unification or of gauge mediation.

The MSSM scalars obtain their supersymmetry-breaking masses from one-loop diagrams
feeding off the gaugino masses and also off the soft mass, $m_\Phi$, of the heavy scalar in $\Phi_{-}$. We expect these masses to be of the order of
\beq
m_q^2 \sim \, \frac{g_i^2}{16 \pi^2} m_i^2\,,\quad\frac{g_i^2}{16 \pi^2} m_\Phi^2\, .
\eqn{scalar}
\eeq
Note that there is also a logarithm corresponding to the usual MSSM running of the scalar masses. In our model this log is cut off at the mass of the heavy gauge bosons $\sim M_i$ so that the log-enhanced contributions are not much larger than the log-less threshold contributions. We discuss details of the computation in the next Section.

From \Eq{scalar}, the scalar masses are expected to be
smaller than gaugino masses by a factor of $4\pi$. However, once we include color factors and the logarithm this factor is reduced to ``a few". Still, scalar superpartners are significantly lighter than gauginos. In addition, colored superpartners are heavier than non-colored ones.

The free parameters of our model which are relevant to the superpartner mass spectrum are
the symmetry-breaking scale $f$ which sets the scale of the heavy gauge multiplets, the soft terms $m_B$ and $m_\Phi^2$, the gauge coupling $g_B$ or equivalently the ratio of gauge couplings $g_{A_3}/g_B\equiv s_3/c_3$, as well as $\mu$ and $B_\mu$. As is customary, we will trade the parameters $\mu$ and $B_\mu$ for the Higgs VEV $v=246$ GeV and $\tan \beta$.
Then the parameter space of the model is
\beq
f\,, \,\, m_B\,, \,\, m^2_\Phi\,, \,\, s_3/c_3\,, \,\, \tan \beta \, .
\eeq

\section{MSSM spectrum}
\label{spectrum}

The MSSM gauginos correspond to the lightest eigenstates of the mass matrix \refeq{gauginomatrix}. Their masses arise at tree-level from the mixing of the $SU(5)_A$ and $SU(5)_B$ gauginos. If the supersymmetry-breaking soft masses are small compared to the masses of the $SU(5)$ gauge multiplets then the gauginos satisfy the usual GUT relations
\beq
\frac{m_3}{\alpha_3}=\frac{m_2}{\alpha_2}=\frac{m_1}{\alpha_1}=\frac{m_B}{\alpha_B}
\,.
\eeq

The MSSM scalar masses in our model are generated from one-loop diagrams involving the gauge multiplets and the link fields, see Fig.~\ref{susy1loop}. In the supersymmetric limit the contributions to scalar masses from all five diagrams must cancel. Including the soft supersymmetry breaking in the gaugino masses $m_B$ and the link field masses $m_\Phi^2$ one obtains {\it finite} scalar masses.  To understand that contributions to scalar masses are not  ultraviolet (UV) divergent note that in the UV the MSSM fields only couple to the A gauginos, which do not have supersymmetry breaking masses. In order to couple to supersymmetry breaking, any one-loop diagram must involve at least two insertions of the mass scale $f$ in addition to the soft supersymmetry-breaking masses, $m_B$ or $m_\Phi^2$. Power counting then tells us that this contribution is UV finite.

\FIGURE[t]{
\includegraphics[scale=1.2]{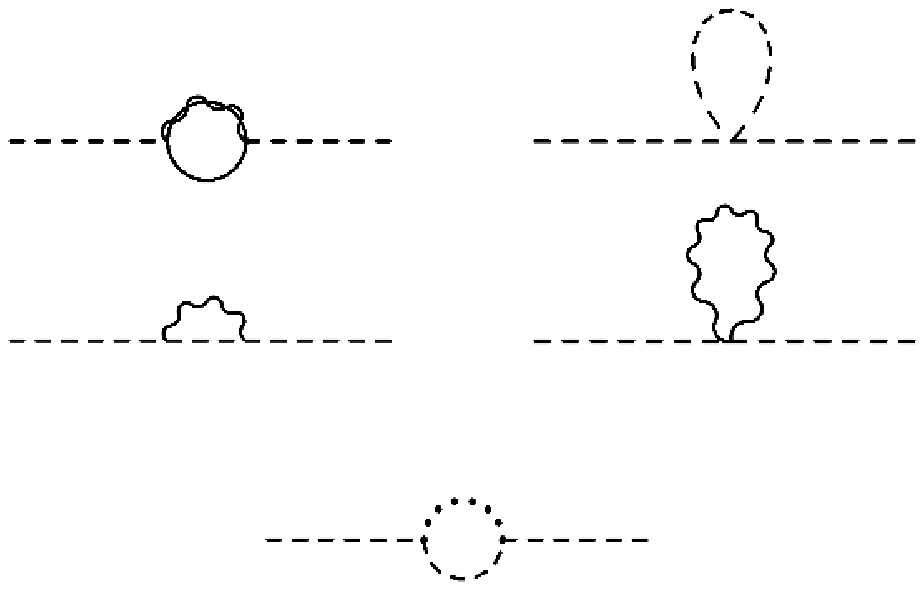}
\caption{One-loop diagrams contributing to the MSSM soft scalar masses.  Dashed, solid and wavy-solid lines  denote scalar, matter fermion and gaugino propagators, respectively. The dotted line represents the real part of the scalar
 in the link field $\Phi_-$\,.  }
\label{susy1loop}
}

To compute the MSSM scalar masses, we can explore the fact that the diagrams in Fig.~\ref{susy1loop} cancel against each other in the supersymmetric limit. This allows us to focus only on the two diagrams  in Fig.~\ref{1loop} and extract the contributions proportional to supersymmetry breaking. In practice, we compute the two diagrams in Fig.~\ref{1loop} and subtract the value of the diagrams with supersymmetry breaking set to zero.

\FIGURE[t]{
 \includegraphics[scale=1.2]{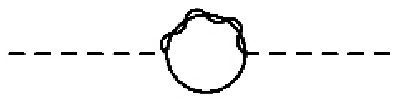}
 \hspace{1.5cm}
 \includegraphics[scale=1.2]{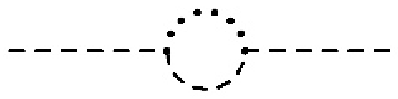}
\caption{One-loop diagrams giving a contribution to the MSSM soft scalar masses proportional
to supersymmetry breaking.  }
\label{1loop}
}

The gaugino diagram in Fig.~\ref{1loop} with gauginos corresponding to the
$i$-th MSSM gauge group and masses $\Omega_i$ (minus the diagram with
supersymmetry preserving gaugino masses $\hat \Omega_i$, defined as in \refeq{gauginomatrix} but with $m_B=0$) gives the scalar mass contribution
\beq
\left.\delta m^2_i\right|_{\textrm{\scriptsize{gaugino}}}=-i{2}g_{A_i}^2 C_2(R_i) \left[
 \int \frac{d^4 k}{(2\pi)^4} \frac{2k^2}{k^2 (k^2-\Omega_i^2)}
 -\int \frac{d^4 k}{(2\pi)^4} \frac{2k^2}{k^2 (k^2-\hat\Omega_i^2)}
 \right]_{11}\, .
\label{gaugino1loop}
\eeq
Here the Casimirs $C_2(R_i)$ depend on the representations $R_i$ of the scalars in question, we give their values
below. The $2k^2$ in the numerator comes from a trace over 2-component spinor indices and we have ignored the standard model fermion mass in the propagator.
The gaugino propagators are $3\times 3$ matrices due to the mass mixing in $\Omega_i$. The MSSM fermions only couple to the A gauginos, thus we really only need the $1-1$ component of the gaugino propagator. Alternatively, we can compute the diagram with the full $3\times 3$ gaugino propagator. Then we obtain a $3\times 3$ matrix as the value of the diagram, and we project onto the $1-1$ component at the end of the calculation. This is the approach we followed in writing Eq.~(\ref{gaugino1loop}), where the notation $[\cdots]_{11}$ stands for the 1-1 component of the matrix in square brackets.
More details about this calculation are given in the Appendix.

In addition to the gaugino diagram there is also the scalar loop in Fig.~\ref{1loop}. The vertices of this diagram originate from the D-terms of the A gauge groups after inserting the  symmetry-breaking VEV $f$. The Feynman rule for this vertex is $-i \sqrt{2} g_{A_i}^2 f\, T^a=-i g_{A_i} s_i M_i T^a$ where $T^a$ is a gauge generator in the fundamental representation. Again, subtracting the supersymmetry preserving diagram, the scalar diagram gives
\beq
\left.\delta m^2_i\right|_{\textrm{\scriptsize{scalar}}}=i g_{A_i}^2 C_2(R_i) s_i^2 M_i^2\! \int\!\! \frac{d^4 k}{(2\pi)^4}\,  \left(  \frac{1}{k^2}\,\frac1{k^2-(M_i^2+2m_\Phi^2)}
-\frac{1}{k^2}\,\frac1{k^2-M_i^2} \right) \ ,
\label{scalar1loop}
\eeq
where $M_i^2+2m_\Phi^2$ is the mass squared of the real part of link field component $\Phi_{-}$ which runs in the loop.

Evaluating the loop integrals in dimensional regularization we obtain for the sum of Eqs.~(\ref{gaugino1loop}) and (\ref{scalar1loop})
\beq
\delta m^2_i
   = C_2(R_i) \left( -\frac{g_{A_i}^2}{4\pi^2}\left[\Omega_i^2 \log \frac{\Omega_i^2}{M_i^2}  \right]_{11}
              + \frac{g_i^2}{16 \pi^2}\frac{s_i^2}{c_i^2} M_i^2\log \frac{M_i^2+2m_\Phi^2}{M_i^2}\right)
              \equiv C_2(R_i) K_i \ .
\eeq
In the limit of small supersymmetry breaking both terms can be expanded in powers of supersymmetry breaking over the heavy gauge boson masses.  We find
\beq
K_i  &\simeq&
 \frac{g_i^2}{4\pi^2} \left(m_{i}^2 \left[\log\left(\frac{M_i^2}{m_i^2}\right)-1+\frac12 \frac{c_i^2}{s_i^2}\right]+m_\Phi^2\left[\frac12\frac{s_i^2}{c_i^2} \right] \right)    \, .
\eqn{oneloopsmallsusy}
\eeq

To get the full one-loop MSSM scalar soft masses we must sum over contributions corresponding to the different gauge quantum numbers. The Casimir $C_2$ for an $SU(N)$ fundamental is $(N^2-1)/2N$, whereas for the $U(1)$ it is $3/5$ times the hypercharge squared. Then
\beq
m_Q^2&=&  \frac43 K_3 + \frac34 K_2 + \frac{1}{60} K_1, \nonumber\\
m_U^2&=&  \frac43 K_3 + \frac{4}{15} K_1,  \nonumber\\
m_D^2&=&  \frac43 K_3 + \frac{1}{15} K_1,  \nonumber\\
m_L^2&=&  \frac34 K_2 + \frac{3}{20} K_1, \nonumber\\
m_E^2&=&   \frac35 K_1,  \nonumber\\
m_{H_u}^2&=&  m_{H_d}^2 = \frac34 K_2 + \frac{3}{20} K_1\ .
\eqn{scalarmasses}
\eeq
Since the Higgs soft masses squared are positive it seems that our model might not break electroweak symmetry. However, because of the large top-Yukawa coupling and because $m_3>m_2$, the negative  two-loop stop-gluino contribution to the $H_u$ soft mass dominates over the one-loop wino and bino contributions computed above.
Our two-loop computation is summarized in Appendix \ref{appendixloop} and gives
\beq
\left. m_{H_u}^2\right|_{\textrm{\scriptsize{2--loop}}}&=&\frac{g_{A_3}^2 \lambda_t^2}{16 \pi^4}\left( \left[ \Omega_3^2 \log^2 \frac{\Omega_3^2}{M_3^2}\right]_{11}
+2 \left(\log \frac {M_3^2}{m_{\tilde t}^2}+1\right) \left[ \Omega_3^2 \log \frac{\Omega_3^2}{M_3^2} \right]_{11}   \right. \nonumber   \\
                      &&\quad\quad\quad -\left.\frac12 s_3^2 M_3^2 \log\left(1+\frac{2 m_\Phi^2}{M_3^2}\right)
                           \left(\frac12\log\left(1+\frac{2 m_\Phi^2}{M_3^2}\right)+ \log\frac{M_3^2}{ m_{\tilde t}^2}+1 \right)\right) \nonumber\\
                               &\simeq&
-\frac{g_3^2 \lambda_t^2}{16 \pi^4}m_{3}^2 \left[
          \log^2 \frac{M_3^2}{m_3^2}+\frac{c_3^2}{s_3^2}\log \frac{M_3^2}{m_3^2}
          -2
        + 2 \log \frac{m_3^2}{m_{\tilde t}^2} \left(\log \frac{M_3^2}{\ m^2_3} \!-\!1\! +\!\frac12 \frac{c_3^2}{s_3^2} \right)\!
        \right]\nonumber \\
        &&-\frac{g_3^2 \lambda_t^2}{16 \pi^4}m_{\Phi}^2 \left[
        \frac{s_3^2}{c_3^2}\left(\log\frac{M_3^2}{m_{\tilde t}^2}+1 \right)\right] \ ,
\eeq
where the second equality holds in the limit of small supersymmetry breaking.

Numerically, the absolute value of this contribution is larger  than the 1-loop result
\beq
\left. m_{H_u}^2\right|_{\textrm{\scriptsize{1--loop}}}&=&\frac34 K_2 + \frac{3}{20} K_1
\nonumber \\
              &\simeq& \frac{g_2^2}{16\pi^2}\left[3 m_2^2 \left(\log \frac{M_2^2}{m^2_2} -1 + \frac12 \frac{c_2^2}{s_2^2} \right)
              + \frac32 \frac{s_2^2}{c_2^2} m_\Phi^2\right]\nonumber \\
              &+&\frac{3}{5}\frac{g_{1}^2}{16\pi^2}\left[m_1^2\left(\log \frac{M_1^2}{m_1^2} -1 + \frac12 \frac{c_1^2}{s_1^2} \right)+\frac12 \frac{s_1^2}{c_1^2} m_\Phi^2\right]\,,
\eeq
so that the full soft mass for $H_u$ is negative, which allows for electroweak symmetry breaking.

Gaugino loops also give rise to A-terms which can be significant for the third generation. They can be written in terms of the matrix elements
\beq
L_i = \frac{g_{A_i}^2}{4 \pi^2} \left[\Omega_i \log \frac{\Omega_i^2}{M_i^2}\right]_{11}
\, ,
\eeq
and we find
\beq
A_{t}&=&\lambda_t \left(\frac43 L_3 + \frac34 L_2 + \frac{13}{60} L_1\right) \\
&\simeq&
\frac{-\lambda_t}{16 \pi^2}\left[
   \frac{32}{3} g_3^2 m_3 \left(\log \frac{M_3}{m_3} - \frac12\right)
   +  6 g_2^2 m_2 \left(\log \frac{M_2}{m_2} - \frac12\right)
   + \frac{26}{15} g_1^2 m_1 \left(\log \frac{M_1}{m_1} - \frac12\right) \right] ,  \cr
A_{b}&=&\lambda_b \left(\frac43 L_3 + \frac34 L_2 + \frac{7}{60} L_1\right) \\
&\simeq&\frac{-\lambda_b}{16 \pi^2}\left[
   \frac{32}{3} g_3^2 m_3 \left(\log \frac{M_3}{m_3} - \frac12\right)
   +  6 g_2^2 m_2 \left(\log \frac{M_2}{m_2} - \frac12\right)
   + \frac{14}{15} g_1^2 m_1 \left(\log \frac{M_1}{m_1} - \frac12\right) \right] , \cr
A_{\tau}&=&\lambda_{\tau} \left(\frac34 L_2 + \frac{9}{20} L_1\right) \\
&\simeq&\frac{-\lambda_{\tau}}{16 \pi^2}\left[
    6 g_2^2 m_2 \left(\log \frac{M_2}{m_2} - \frac12\right)
   + \frac{18}{5} g_1^2 m_1 \left(\log \frac{M_1}{m_1} - \frac12\right) \right]\nonumber\,.
\eeq

The Higgs mass in this model gets a new small contribution which is absent in the MSSM. It stems from the D-terms of the heavy gauge bosons which do not completely decouple if there is supersymmetry breaking. By integrating out the heavy link scalar field at tree level we derive that the usual MSSM D-terms for the Higgs doublets are modified to
\beq
V_D=\frac{g_2^2(1+\Delta_2)}{8} \left|H_u^\dagger \sigma^a H_u +H_d^\dagger \sigma^a H_d\right|^2
   +\frac{\frac35g_1^2(1+\Delta_1)}{8} \left|H_u^\dagger H_u -H_d^\dagger  H_d\right|^2\,,
\eqn{quartic}
\eeq
where $\sigma^a$ are the Pauli matrices, and the $\Delta_i$ are given by
\beq
\Delta_i&=&\frac{s_i^2}{c_i^2} \, \frac{2m_\Phi^2}{M_i^2+2m_\Phi^2} \ .
\eeq
These corrections are positive when the link field soft mass is positive and negative otherwise.
They vanish in the limit $m_\Phi^2 \rightarrow 0$ as they should by supersymmetry.
For our example spectrum these corrections turn out to be small and increase the Higgs mass by only 1\%.

\TABLE{
\begin{tabular}{c|c|c}
& & mass\\
\hline
inputs: & $f$ & $5000$  \\
 & $m_{B}$ & $5000$ \\
 & $m_{\tilde{\Phi}}$ & $5000$ \\
 & $\tan\beta$ & 8 \\
 & $g_{A_3}/g_B$ & $0.8$  \\
 \hline heavy gauge bosons: & $M_3$ &15400 \\
 & $M_2$ & 12970 \\
 & $M_1$ & 12500\\
 \hline gluino: & $m_3$ & 1904 \\
 \hline neutralinos: & $m_{\chi^0_1}$ & 232 \\
 & $m_{\chi^0_2}$ & 253 \\
 & $m_{\chi^0_3}$ & 383 \\
 & $m_{\chi^0_4}$ & 706 \\
 \hline charginos: & $m_{\chi^\pm_1}$ & 243 \\
 & $m_{\chi^\pm_2}$ & 706  \\
 \hline Higgs:
 & $m_{h^0}$ & 116 \\
 & $m_{H^0}$ & 324 \\
 & $m_A$ & 324 \\
 & $m_{H^\pm}$ & 334 \\
 & $\mu$ & 249  \\
 & $\sqrt{B_\mu}$ & 114 \\
 \hline sleptons: & $m_{\tilde{e}_R}$ & 102 \\
 & $m_{\tilde{e}_L}$ & 218 \\
 & $m_{\tilde{\nu}_L}$ & 203 \\
 \hline squarks: & $m_{\tilde{u}_L}$ & 934  \\
 & $m_{\tilde{u}_R}$ & 914 \\
 & $m_{\tilde{d}_L}$ & 938  \\
 & $m_{\tilde{d}_R}$ & 913
\end{tabular}
\caption{An example spectrum of low-scale gaugino mediation. All masses are in GeV. Note that the right-handed slepton masses are very near the LEP bounds. Heavier spectra can be obtained by scaling up all input masses.}
\label{tablespectra}
}

More significant is the usual top-stop loop correction to the Higgs potential which is large in this model because we have heavy stops
\beq
\Delta(m_{h^0}^2)=\frac{3}{4\pi^2}\cos^2\! \alpha\, \lambda_t^2 m_t^2 \log\frac{m^2_{\tilde t}}{m^2_t}\,,
\eeq
where $\alpha$ is the angle which diagonalizes the Higgs mass matrix, it becomes identical to $\beta$ in the
decoupling limit, see Ref.~\cite{martin}. With this large correction, we generically
find a Higgs mass above the LEP bound.

To determine the superpartner mass spectrum using the soft masses computed above we use the tree-level formulas summarized in Ref.~\cite{martin}.
Table \ref{tablespectra} gives an example spectrum. It is clear
that gauginos are heavier than scalars and that colored particles
are heavier than non-colored ones. Note that even the left-handed
sleptons are lighter than any charginos or neutralinos. Throughout
parameter space, we expect the gravitino to be the LSP and a
right-handed slepton is the Next-to-LSP (NLSP).

\section{Phenomenology}
\label{pheno}

\subsection{Leptons galore at the LHC}

The most significant feature of our spectrum with regards to collider signatures is that all sleptons are light. Since the NLSP is a right-handed slepton all decay chains end with the production of a lepton which carries the opposite lepton number of the NLSP. Depending on the scale of primordial supersymmetry breaking, the NLSP may either be sufficiently long lived to escape the detectors without decaying or else decay within the detector and lead to displaced vertices. For simplicity of discussion, we will assume that the NLSP is long-lived and appears as a stable charged track, then there is no significant missing energy. This is familiar from gauge mediation and gaugino mediation models in which the right-handed sleptons are lighter than the Bino.

\FIGURE[t]{
 \includegraphics[scale=1.0]{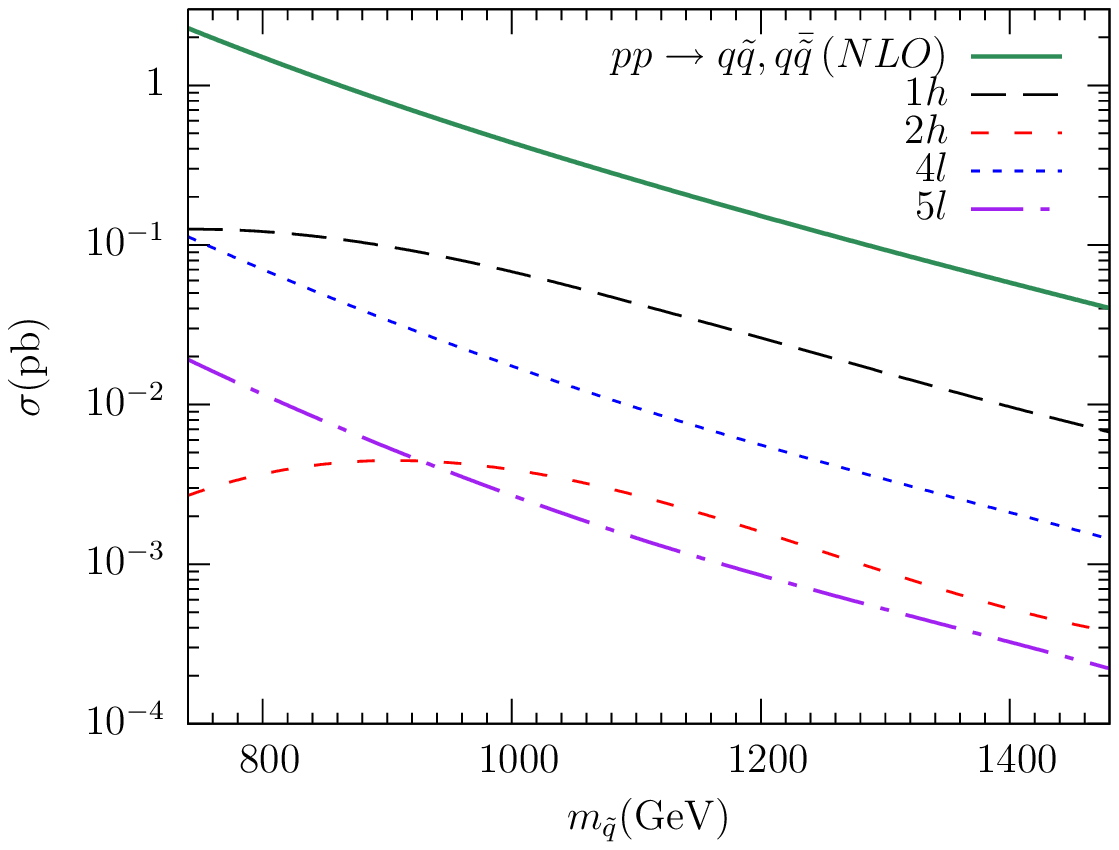}
 \caption{Cross section for squark pair production as a function of squark mass, and cross section times branching fraction for a number of interesting final states with multi-leptons or Higgses. In this plot, we uniformly rescale superpartner masses compared to those in Table 1. The squark mass of 920~GeV corresponds to the spectrum in Table 1. Note that each final state also includes at least two jets and two stable charged tracks. The initial LHC annual integrated luminosity is expected to be $10\, {\rm fb}^{-1}$ and eventually reach $100\, {\rm fb}^{-1}$. Thus a cross section of $10^{-3}$ pb produces 10 events per year already at low luminosity which is sufficient for a discovery in background free channels. }
\label{cross-sections}
}

What is different in our model is that the left-handed sleptons are also lighter than all neutralinos and charginos%
\footnote{This may also occur when there are large negative D-term contributions to left-handed sleptons \cite{Covi}.}  and therefore many decay chains also pass through the left-handed sleptons. This leads to the production of additional leptons. Thus a typical signal for squark pair production at the LHC is not jets and missing energy but instead production of two jets, several leptons, a pair of heavy charged tracks with no significant missing energy.

Consider the decay of a left-handed up-type squark. A possible decay chain is
\beq
\tilde u_L
\arr[55\%] d_L \tilde W^+
\arr[20\%] d_L l_L^+ \tilde \nu
\arr[66\%] d_L l_L^+ \nu l_R^\pm \tilde l_R^\mp\, , \nonumber
\eeq
where $l_L$ and $l_R$ stand for left- and right-handed leptons, respectively. For this simple phenomenological study we only included muons and electrons in the lepton count.
Thus we have two charged leptons, a stable charged track from the slepton, one jet and some missing energy from the neutrino. The total branching fraction for this decay chain is approximately 7\%. Even more striking is the decay into 3 leptons which occurs with a branching fraction of about 3\% via
\beq
\tilde u_L
\arr[28\%] u_L \tilde W^0
\arr[20\%] u_L l_L^\pm \tilde l_L^\mp
\arr[55\%] u_L l_L^\pm  l_L^\mp  (l_R^\pm \tilde l_R^\mp)\, , \nonumber
\eeq
where the parenthesis have been added to indicate the correlation of the signs of the charges. For example, a pair of squarks decays into 4 leptons, 2 jets, and 2 stable charged tracks via the above (and similar) decay chains with a branching fraction of $4\%$, or into 5 leptons, 2 jets and 2 stable charged tracks with a branching fraction of $1\%$. Both final states are background free and very easy to trigger on.

So far we have ignored the fate of the right-handed sleptons in the final state. For the parameter values chosen in our example point, smuons and selectrons can decay to staus via
$\tilde l_R \rightarrow l_R \tau_R^\pm \tilde \tau_R^\mp$. This decay has a long lifetime and might lead to displaced vertices. However, we expect it to be rather difficult to observe because the lepton and $\tau$ produced in the decay are very soft, and the charged track will barely display a kink. For a detailed study see \cite{Ambrosanio:1997bq, Allanach:2001sd}.

In Figure \ref{cross-sections} we display the cross section at the LHC for production of squark pairs as a function of the masses of the squarks. The lower curves show the expected cross section times branching ratios into particular final states using decay chains similar to the ones shown above. This plot may be used to estimate the expected reach in squark masses for any particular channel. For example, for final states with negligible standard model background (such as $\ge 3$ lepton final states) one would look for the mass at which the cross section times integrated luminosity drops below a few events.

Another interesting consequence of the fact that left-handed and right-handed sleptons are at the bottom of the sparticle spectrum is that left-handed slepton decays into Higgs bosons and right-handed sleptons become competitive even though they are Yukawa coupling suppressed. In fact, we find that left-handed staus decay into Higgses 42\% of the time, and even left-handed smuons decay into Higgses when there is enough phase space. For our example point, the $\tilde \mu_L \rightarrow h \tilde \mu_R$ phase space is so small that radiative corrections (which we did not calculate) will determine whether this channel is open (and dominant). For the purposes of the decay chains above we assumed that this decay is closed.
Obviously, the decays into Higgses would be very interesting as they would provide an easy discovery channel for the Higgs. Every such event would be accompanied with stable charged tracks and leptons for easy triggering so that Higgs decay into bottom pairs should be observable.

\subsection{Precision electroweak constraints}

Our model has an extended gauge sector that includes heavy $W'$ and $Z'$
gauge bosons. These bosons mediate interactions which contribute to  the precision electroweak observables,
and therefore there is a lower bound on their masses. The link fields  contain an $SU(2)$ triplet field which couples to the Higgs bosons and
affects the $\rho$ parameter.
It is straightforward to  integrate out the $W'$ and $Z'$ bosons as  well as the scalar triplet at tree level to obtain an effective  description with higher dimensional  operators. The constraints from  experiments on the coefficients of these operators have been  determined in Ref.~\cite{Han:2004az}.
The most important contributions of the heavy gauge bosons are to  four-fermion operators and operators that are  products of fermion  and Higgs currents. Additionally, the $Z'$ and the triplet contribute
to custodial symmetry violation. In the notation of  Ref.~\cite{Han:2004az}
the coefficients of the dimension six operators are
\begin{eqnarray}
 &&
    a_h = \left(\frac{5}{3}\right) \frac{1}{4}  \frac{g_Y^4}{f^2  g_B^4} - \frac{2 f^2 g_{A_2}^4 }{M_\phi^4} \cos^2(2 \beta), \ \ \
    a^s_{h\psi}=  \left(\frac{5}{3}\right) \frac{1}{4}  Y_\psi \frac {g_Y^4}{f^2 g_B^4},  \nonumber  \\
&&
    a^s_{\psi \psi'} = \left(\frac{5}{3}\right) \frac{1}{2}  Y_\psi  Y_\psi'  \frac{g_Y^4}{f^2 g_B^4},  \ \ \
    a^t_{\psi \psi'} = - \frac{1}{8} \frac{g_2^4}{f^2 g_B^4} , \ \ \
    a^t_{h \psi} = \frac{1}{8} \frac{g_2^4}{f^2 g_B^4}  , \ \ \  \nonumber
\end{eqnarray}
where $Y_\psi$ is the hypercharge of fermion $\psi$ and the factors  of $\frac{5}{3}$ come
from converting $SU(5)$ normalized hypercharge couplings to SM  normalization. The triplet contribution is proportional to $\cos^2(2  \beta)$,
where $M_\phi^2=2 f^2 (g_{A_2}^2+ g_B^2) + 2 m_\Phi^2$ is the mass of  the scalar triplet.

Note that the operator  coefficients induced by the heavy gauge  bosons depend on the free parameters of
our model only in the combination $f^2 g_B^4$. The contribution from  the link field triplet field
is more complicated as it depends on the soft mass of the link field  and $\tan\beta$.

We could find a simultaneous bound on the contributions of the
gauge  bosons and the scalar triplet, but to be conservative we
provide independent bounds on the two contributions (which
partially cancel). We neglect the triplet contribution when
constraining the gauge bosons and conversely  neglect gauge bosons
when constraining the triplet. At $3 \sigma$ confidence level we
obtain the bounds $f g_B^2 \geq 1.8\ {\rm TeV}$ and $M_\Phi^2/( f
g_{A_2}^2) \geq  7.8\ {\rm TeV}$ assuming $\cos(2 \beta) \approx
1$. These bounds are not very stringent given the mass scales of
interest  to us. For the spectrum in Table 1, the heavy gauge
bosons and  the triplet  fields are much heavier then the
corresponding electroweak constraints ($f g_B^2 \simeq 14\ $TeV
and $M_\Phi^2/ (f g_{A_2}^2) \simeq 176 \ $TeV).

\subsection{Cosmological constraints}

Although a detailed analysis of all the cosmological constraints and  aspects of our model is beyond the scope of the present work, we briefly discuss some of the salient points.

In our model the gravitino is the
LSP and a right-handed slepton is the NLSP.
If $R$-parity is conserved, the LSP is stable and it may be the dark matter.%
\footnote{The situation for dark matter is very similar to that in
gauge mediation. Alternative dark matter candidates other than the gravitino
 in theories with gauge-and gaugino-mediated supersymmetry breaking have been studied in
 Ref.~\cite{dgp}. }
The major cosmological constraints concerning the LSP+NLSP system  are: avoiding that the LSP energy density  exceeds the critical density of the universe, and avoiding that the decay products of the NLSP upset the successful predictions of the Big Bang Nucleosynthesis (BBN) for
the light element abundances.

The requirement that the energy density of  thermal gravitinos does not overclose the universe leads  to  a bound on the gravitino mass $m_{3/2}\lesssim$ keV \cite{gravitinomass}.
Gravitinos in the keV range may serve as a warm dark matter component of the universe;
for the typical mass spectra of our model, such a possibility seems to be
disfavored by the
analysis of Ref.~\cite{gorbunov},
due to a significant fine-tuning of the reheating temperature and a narrow parameter space.

On the other hand, if $m_{3/2}\gtrsim$  keV, for which the gravitino would be a cold dark matter candidate, some mechanism of large  entropy production (e.g.~inflation or out-of-eq\-uilibrium decay of a heavy particle) is needed to dilute the gravitino abundance  down to an acceptable level and to avoid overclosure of the universe.
With such a  mechanism  at work, any pre-existing gravitino abundance gets
depleted. However, a population of gravitinos is regenerated in the primordial plasma
in two ways: by thermal scattering, whose effectiveness increases with the temperature,
and by NLSP decays.
  Therefore,  the energy density constraint on the total abundance of gravitinos puts an upper bound on the temperature $T_R$  at which the dilution mechanism ends and the  ordinary radiation-dominated epoch begins \cite{MMY}.
In the case where inflation is the main entropy-releasing process in the early universe, $T_R$ corresponds to the reheating temperature.

The BBN constraint mostly concerns the lifetime of the NLSP,
$\tau_{\textrm{\scriptsize{NLSP}}}$.
Recent analyses \cite{cbbn}
have shown that successful BBN requires
$\tau_{\textrm{\scriptsize{NLSP}}}\lesssim 5\times 10^3$ s.
 For the typical
masses in our model, $m_{\textrm{\scriptsize{NLSP}}}\sim 100$ GeV
and $m_{3}\sim 1$ TeV,
the combination of the overclosure and BBN constraints translates into
 upper bounds on the gravitino mass, $m_{3/2}\lesssim 1$ GeV, and
 on the reheating temperature, $T_R\lesssim 10^7$ GeV.
Scenarios where a substantial amount of entropy is produced after inflation by
the out-of-equilibrium decay of a heavy particle,
like e.g.~the one in Ref.~\cite{FY} where the lightest messengers
decay at late times, would allow even higher reheating
temperatures.

\section{Future directions}
\label{conclusions}

We see several avenues for interesting further work on model building and the phenomenology of low-scale gaugino mediation. With regards to phenomenology, it seems clear that these models offer signatures which should be explored further. In addition to the signatures with many leptons there are also interesting decays into the relatively light Higgs/Higgsino sector as well as into the Higgs. Because the NLSP is visible, superpartner masses should be easy to reconstruct via bumps in invariant mass distributions. Clearly, a more complete and precise phenomenological study of the different signatures and the reach would be interesting.

The spectrum of low-scale gaugino mediation is similar to the spectrum of the MSSM with Dirac gaugino masses~\cite{Fox:2002bu}.
It should be possible to distinguish between our Majorana gaugino masses and Dirac gaugino masses via charge asymmetries between the leptons in final states. Another model which gives a similar spectrum is low-scale gauge mediation with a very large number of messengers, where gaugino masses are enhanced relative to scalar masses by a factor equal to the square root of the number of messengers.

With regards to model building, it would be interesting to do away with the explicit scale $\Lambda$ in the superpotential and generate the link field VEVs from a dynamical mechanism tied to supersymmetry breaking. For example, if the supersymmetry breaking masses for the adjoint $A$ and the link fields $\Phi$ are comparable at very high scales, then renormalization due to the Yukawa coupling $\lambda_A$ will drive the link field soft mass negative
and naturally generate a link field VEV on the order of the soft masses. What makes this scenario less straightforward is the fact that negative link field soft masses lead to negative contributions to the MSSM scalar masses as can be seen from \refeq{oneloopsmallsusy} and to negative contributions to the Higgs quartic coupling
\refeq{quartic}.

We have focused primarily on the spectrum of low-scale gaugino mediation and allowed ourselves to choose $\mu$ and $B_\mu$  to give successful electroweak symmetry breaking. A more complete model should involve a mechanism for generating $\mu$ and $B_\mu$ with acceptable size. Presumably this would involve a modification of the Higgs sector so that $H_u$ and $H_d$ can couple more directly to supersymmetry breaking. Another issue is gauge coupling unification. Naively, the contributions of the link fields to the running of the $SU(3) \times SU(2) \times U(1)$ gauge couplings preserve unification because they come in complete $SU(5)$ multiplets. However, they make too large a contribution to the overall running so that the gauge couplings become large before unifying.

\acknowledgments

We thank Andy Cohen, Lil' Dave, Ambar Jain, Tilman Plehn, Veronica Sanz,  and Ignazio Scimemi for valuable discussions.
We also thank Tilman for computing LHC production cross sections in our model.
MS and WS thank the Aspen Center for Physics and the Santa Barbara KITP for their hospitality during the completion of this work.
The work of ADS is supported in part by the INFN ``Bruno Rossi'' Fellowship,
 and in part by the U.S. Department of Energy (DoE) under contract No.
 DE-FG02-05ER41360. J.F  and W.S. are supported in part by the DoE under grant No. DE-FG-02-92ER40704, M.S. is supported by grant No. DE-FG02-01ER40676.

\appendix

\section{The loop computations}
\label{appendixloop}

The gaugino contribution to one-loop scalar masses are given by Eq.~(\ref{gaugino1loop}) where $\Omega_i$ and $\hat \Omega_i$ are matrices. To avoid notational clutter while computing the integral, we drop the index ``i". We compute the
integral involving $\Omega$ and $\hat \Omega$ separately. First, we diagonalize $\Omega$ by writing
$\Omega = U^\dagger D U$ where $U$ is a unitary matrix and $D$ is diagonal. Then the integral in Eq.~(\ref{gaugino1loop}) may be written as
\beq
I[\Omega]=\left[ U^\dagger
 \int \frac{d^4 k}{(2\pi)^4} \frac{2k^2}{k^2 (k^2-D^2)}\ U
 \right]_{11} \, .
 \eeq
This is easily evaluated in dimensional regularization as $D$ is diagonal. We find
\beq
I[\Omega]=\frac{-2i}{16\pi^2}
\left[U^\dagger D^2 \left(-\frac{1}{\bar \epsilon} -1 + \log \frac{D^2}{\mu^2}\right)U \right]_{11}
=\frac{-2i}{16\pi^2}
\left[\Omega^2 \left(-\frac{1}{\bar \epsilon} -1 + \log \frac{\Omega^2}{\mu^2}\right)\right]_{11}\! .
\eeq
When we subtract the corresponding expression for $\hat \Omega$ the divergent terms, the $1$, and the $\log \mu^2$ terms cancel because the 1-1 elements of $\Omega^2$ and $\hat \Omega^2$ are identical. Thus
we may write
\beq
I[\Omega]-I[\hat \Omega] = \frac{-2i}{16\pi^2}
\left[\Omega^2 \log \frac{\Omega^2}{M^2}-
\ \hat \Omega^2 \log \frac{\hat \Omega^2}{M^2}\right]_{11} \ ,
\eeq
where $M$ is an arbitrary mass scale which we choose equal to the heavy gauge boson mass.
But with $M$ equal to the heavy gauge boson mass, the second term vanishes as can be seen by diagonalizing
$\hat \Omega^2$
\beq
\hat \Omega^2 \log \frac{\hat \Omega^2}{M^2} =
   M^2 \hat U^\dagger \pmatrix{0&0&0 \cr 0&1&0\cr 0& 0&1}
\log \pmatrix{0&0&0 \cr 0&1&0\cr 0& 0&1} \hat U = 0 \ .
\eeq
Thus we now have
\beq
\left.\delta m^2_i\right|_{\textrm{\scriptsize{gaugino}}}=
-\frac{g_{A_i}^2}{4 \pi^2} C_2(R_i)\left[\Omega_i^2 \log \frac{\Omega_i^2}{M_i^2}\right]_{11}  \ .
\eeq
This may either be evaluated by numerically diagonalizing $\Omega_i$ or by expanding in small supersymmetry breaking. In the latter case we find
\beq
g^2_{A_i} \left[\Omega_i^2 \log \frac{\Omega_i^2}{M_i^2}\right]_{11} \simeq
-g_i^2 m_i^2 \left(\log \frac{M_i^2}{m_i^2} -1 + \frac12 \frac{c_i^2}{s_i^2} \right) \ .
\eeq

The link field scalar contribution is easy to evaluate directly in dimensional regularization. The result is
\beq
\left.\delta m^2_i\right|_{\textrm{\scriptsize{scalar}}}=\frac{g_i^2}{16\pi^2}  \frac{s_i^2}{c_i^2} C_2(R_i) M_i^2
\log \frac{M_i^2 + 2 m_\Phi^2 }{M_i^2}\simeq
\frac{g_i^2}{8\pi^2}  \frac{s_i^2}{c_i^2} C_2(R_i)\, m_\Phi^2\ \ .
\eeq

The computation of the A-terms is completely analogous.

To evaluate the two-loop diagrams contributing the $H_u$ soft mass we proceed similarly. Again, we compute the diagrams involving supersymmetry breaking and subtract their supersymmetry preserving pieces. Also, for simplicity, we only include diagrams proportional to the largest couplings ($\lambda_t, g_{A_3}$) so that our final answer is porportional to $\lambda_t^2 g_3^2$.

Special care needs to be taken to avoid infrared divergences from the massless stop propagators in these diagrams. To deal with this IR divergence we add and subtract the stop mass to the Lagrangian. We treat the added stop mass exactly by working with massive stop propagators and we treat the subtracted stop mass perturbatively as a mass insertion counterterm.
Since the stop mass is already of order $g^2$, we only need to insert this counterterm in one-loop diagrams to obtain a consistent result to order $g^2 \lambda_t^2$. Then there are 5 relevant diagrams to compute, see Fig.~\ref{2loop}.

\FIGURE[t]{
\begin{tabular}{ccc}
 \includegraphics[scale=0.5]{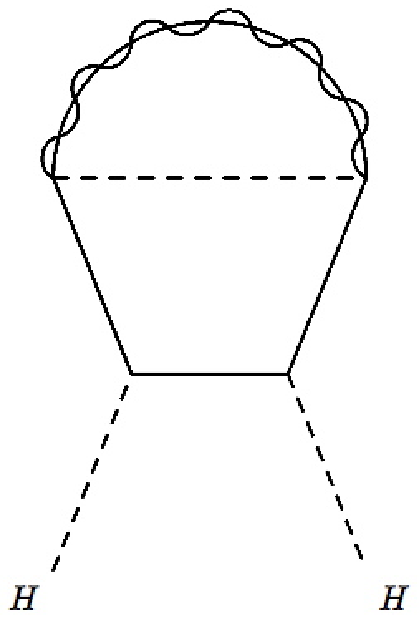}(I)&
\includegraphics[scale=0.47]{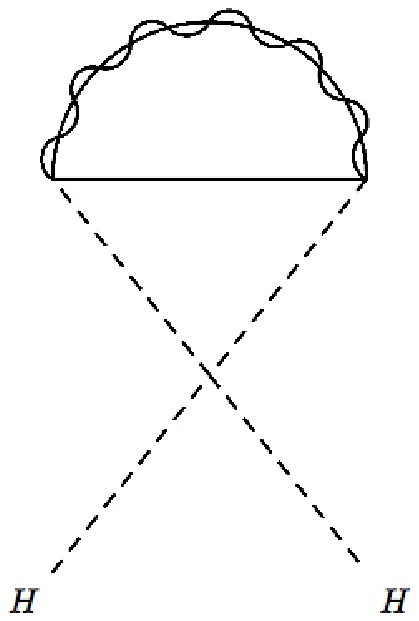}(II)&
\includegraphics[scale=0.47]{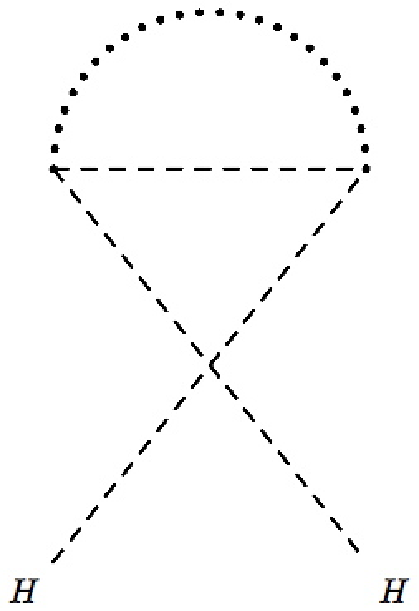}(III)\\
\includegraphics[scale=0.5]{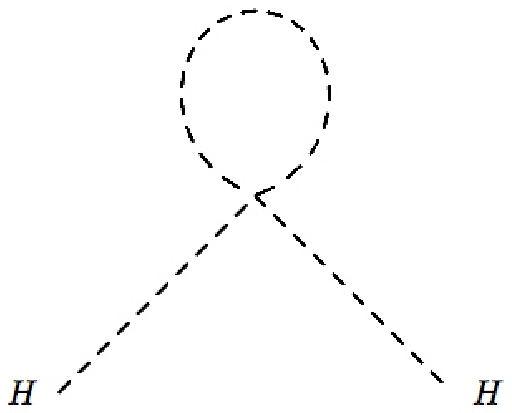}(IV)&&
\includegraphics[scale=0.5]{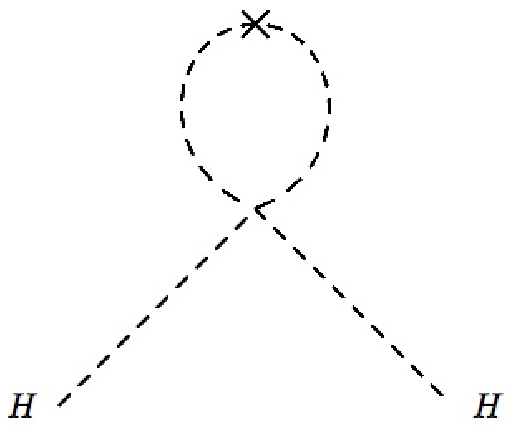}(V)
\end{tabular}
\caption{Diagrams contributing to the Higgs self-energy, to order $g^2\lambda_t^2$. The Higgs field is on the external legs. Internal dashed, solid and wavy-solid lines  denote stop, top and gaugino propagators, respectively. The dotted line represents the real part of the link field component $\Phi_-$\,. }
\label{2loop}
}

After subtracting the supersymmetry-preserving parts the 5 diagrams can be written as
\beq
m^2_{\textrm{\scriptsize{\,I}}}&=&\frac{g_{A_3}^2 \lambda_t^2}{16 \pi^4}\left(2 \left[ \Omega_3^2 \log \frac{\Omega_3^2}{M_3^2}\right]_{11}\right)\, ,\\
m^2_{\textrm{\scriptsize{\,II}}}&=&\frac{g_{A_3}^2 \lambda_t^2}{16 \pi^4}\left( \left[ \Omega_3^2 \log^2 \frac{\Omega_3^2}{m_{\tilde t}^2}-\hat \Omega_3^2 \log^2 \frac{\hat \Omega_3^2}{m_{\tilde t}^2}\right]_{11}
                       -2 \left[ \Omega_3^2 \log \frac{\Omega_3^2}{M_3^2}\right]_{11}\right)
                    \, ,   \\
m^2_{\textrm{\scriptsize{\,III}}}&=&-\frac{g_3^2 \lambda_t^2}{64 \pi^4} \frac{s_3^2}{c_3^2}
                       M_3^2 \log \left(1+\frac{2m_\Phi^2}{M_3^2}\right)
                       \left(\log \left(1+\frac{2m_\Phi^2}{M_3^2}\right)+2 \log\frac{M_3^2}{m_{\tilde t}^2}\right)\ ,\\
m^2_{\textrm{\scriptsize{\,IV}}}& =&-\frac{6 \lambda_t^2}{16 \pi^2} m_{\tilde t}^2
      \left(\frac{1}{\bar \epsilon} -\log\frac{m_{\tilde t}^2}{\mu^2} +1\right)\, , \\
m^2_{\textrm{\scriptsize{\,V}}}& =&\ \ \,\frac{6 \lambda_t^2}{16 \pi^2} m_{\tilde t}^2
      \left(\frac{1}{\bar \epsilon} -\log\frac{m_{\tilde t}^2}{\mu^2} \right)\, ,
\eeq
where we have dropped anything that is proportional to $[\Omega_3^2]_{11}$ without any $\log \Omega_3^2$ because it vanishes after subtracting the supersymmetry-preserving pieces.
To evaluate the first term in $m^2_{\textrm{\scriptsize{\,II}}}$ we write
\beq
\left[ \Omega_3^2 \log^2 \frac{\Omega_3^2}{m_{\tilde t}^2}- (\Omega_3 \leftrightarrow \hat \Omega_3) \right]_{11} &=&
\left[ \Omega_3^2 (\log \frac{\Omega_3^2}{M_3^2}+\log \frac {M_3^2}{m_{\tilde t}^2})^2 -(\Omega_3 \leftrightarrow \hat \Omega_3) \right]_{11}\\
& =&\left[ \Omega_3^2 (\log^2 \frac{\Omega_3^2}{M_3^2}+2 \log \frac{\Omega_3^2}{M_3^2} \log \frac {M_3^2}{m_{\tilde t}^2}) -(\Omega_3 \leftrightarrow \hat \Omega_3) \right]_{11}\\
&=&\left[ \Omega_3^2 \log^2 \frac{\Omega_3^2}{M_3^2}\right]_{11}
+2 \log \frac {M_3^2}{m_{\tilde t}^2} \left[ \Omega_3^2 \log \frac{\Omega_3^2}{M_3^2} \right]_{11}\! .
\eeq

Adding the 5 diagrams we have
\beq
\!\!\!\!\!\!
 \left. m^2_{H_u}\right|_{\textrm{\scriptsize{2--loop}}}&=&\frac{g_{A_3}^2 \lambda_t^2}{16 \pi^4}\left\{
 \left[ \Omega_3^2 \log^2 \frac{\Omega_3^2}{M_3^2}\right]_{11}
+2 \left(\log \frac {M_3^2}{m_{\tilde t}^2}+1\right) \left[ \Omega_3^2 \log \frac{\Omega_3^2}{M_3^2} \right]_{11}                       \right. \nonumber   \\
                      &&\quad\, -\left.\frac12 s_3^2 M_3^2 \log\left(1+\frac{2 m_\Phi^2}{M_3^2}\right)
                           \left(\frac12\log\left(1+\frac{2 m_\Phi^2}{M_3^2}\right)+ \log\frac{M_3^2}{ m_{\tilde t}^2}+1 \right)\right\} \! .
\eqn{allfive}
\eeq
It remains to evaluate the matrix elements. As before, we may do this numerically for given values or analytically by expanding in small supersymmetry breaking. For the latter, we need
\beq
g_{A_3}^2 \left[\Omega_3^2 \log^2 \frac{\Omega_3^2}{M_3^2}\right]_{11}&\simeq&
g_3^2 m_3^2 \left(\log^2 \frac{M_3^2}{m_3^2} + \frac{c_3^2}{s_3^2}\right) \\
g_{A_3}^2 s_3^2 \frac{M_3^2}{2} \log\left(1\!+\!\frac{2 m_\Phi^2}{M_3^2}\right) &\simeq& g_3^2 \frac{s_3^2}{c_3^2}\, m_\Phi^2 \ ,
\eeq
so that the sum of all 5 diagrams  becomes
\beq
\left. m_{H_u}^2\right|_{\textrm{\scriptsize{2--loop}}}&\simeq&-\frac{g_3^2 \lambda_t^2}{16 \pi^4}m_3^2 \left[
          \log^2 \frac{M_3^2}{m_3^2}+\frac{c_3^2}{s_3^2}\log \frac{M_3^2}{m_3^2}
          -2
        + 2 \log \frac{m_3^2}{m_{\tilde t}^2} \left(\log \frac{M_3^2}{\ m^2_3} \!-\!1\! +\!\frac12 \frac{c_3^2}{s_3^2} \right)\!
        \right]\nonumber \\
        &&-\frac{g_3^2 \lambda_t^2}{16 \pi^4}m_{\Phi}^2 \left[
        \frac{s_3^2}{c_3^2}\left(\log\frac{M_3^2}{m_{\tilde t}^2}+1 \right)\right] \ .
\eeq


\begin{thebibliography}{99}

\bibitem{Kaplan:1999ac}
  D.~E.~Kaplan, G.~D.~Kribs and M.~Schmaltz,
  Phys.\ Rev.\  D {\bf 62}, 035010 (2000)
  [arXiv:hep-ph/9911293];
  Z.~Chacko, M.~A.~Luty, A.~E.~Nelson and E.~Ponton,
  JHEP {\bf 0001} (2000) 003
  [arXiv:hep-ph/9911323].

\bibitem{Schmaltz:2000gy}
  M.~Schmaltz and W.~Skiba,
  Phys.\ Rev.\  D {\bf 62}, 095005 (2000)
  [arXiv:hep-ph/0001172],
  Phys.\ Rev.\  D {\bf 62}, 095004 (2000)
  [arXiv:hep-ph/0004210];
  H.~Baer, A.~Belyaev, T.~Krupovnickas and X.~Tata,
  Phys.\ Rev.\  D {\bf 65} (2002) 075024
  [arXiv:hep-ph/0110270].

\bibitem{Dine:1994vc}
  M.~Dine, A.~E.~Nelson and Y.~Shirman,
  Phys.\ Rev.\  D {\bf 51}, 1362 (1995)
  [arXiv:hep-ph/9408384];
  M.~Dine, A.~E.~Nelson, Y.~Nir and Y.~Shirman,
  Phys.\ Rev.\  D {\bf 53}, 2658 (1996)
  [arXiv:hep-ph/9507378];
  G.~F.~Giudice and R.~Rattazzi,
  Phys.\ Rept.\  {\bf 322}, 419 (1999)
  [arXiv:hep-ph/9801271].




\bibitem{Cheng:2001an}
  H.~C.~Cheng, D.~E.~Kaplan, M.~Schmaltz and W.~Skiba,
  Phys.\ Lett.\  B {\bf 515}, 395 (2001)
  [arXiv:hep-ph/0106098].

\bibitem{Csaki:2001em}
  C.~Csaki, J.~Erlich, C.~Grojean and G.~D.~Kribs,
  Phys.\ Rev.\  D {\bf 65}, 015003 (2002)
  [arXiv:hep-ph/0106044].

\bibitem{Batra}
  P.~Batra, A.~Delgado, D.~E.~Kaplan and T.~M.~P.~Tait,
  JHEP {\bf 0402}, 043 (2004)
  [arXiv:hep-ph/0309149].


\bibitem{martin}
S.~P.~Martin,
  ``A supersymmetry primer,''
  arXiv:hep-ph/9709356.

\bibitem{Covi}
  L.~Covi and S.~Kraml,
  JHEP {\bf 0708}, 015 (2007)
  [arXiv:hep-ph/0703130];
  S.~Kraml and D.~T.~Nhung,
  JHEP {\bf 0802}, 061 (2008)
  [arXiv:0712.1986 [hep-ph]].

\bibitem{Ambrosanio:1997bq}
  S.~Ambrosanio, G.~D.~Kribs and S.~P.~Martin,
  Nucl.\ Phys.\  B {\bf 516}, 55 (1998)
  [arXiv:hep-ph/9710217].

\bibitem{Allanach:2001sd}
  B.~C.~Allanach, C.~M.~Harris, M.~A.~Parker, P.~Richardson and B.~R.~Webber,
  JHEP {\bf 0108}, 051 (2001)
  [arXiv:hep-ph/0108097].

\bibitem{Han:2004az}
  Z.~Han and W.~Skiba,
  Phys.\ Rev.\  D {\bf 71}, 075009 (2005)
  [arXiv:hep-ph/0412166].

\bibitem{dgp}
S.~Dimopoulos, G.~F.~Giudice and A.~Pomarol,
  Phys.\ Lett.\  B {\bf 389} (1996) 37
  [arXiv:hep-ph/9607225];
  W.~Buchmuller, L.~Covi, J.~Kersten and K.~Schmidt-Hoberg,
  JCAP {\bf 0611}, 007 (2006)
  [arXiv:hep-ph/0609142].

\bibitem{gravitinomass}
  H.~Pagels and J.~R.~Primack,
  Phys.\ Rev.\ Lett.\  {\bf 48} (1982) 223.

\bibitem{gorbunov}
  D.~Gorbunov, A.~Khmelnitsky and V.~Rubakov,
  arXiv:0805.2836 [hep-ph].

\bibitem{MMY}
  T.~Moroi, H.~Murayama and M.~Yamaguchi,
  Phys.\ Lett.\  B {\bf 303} (1993) 289.

 \bibitem{cbbn}
   M.~Pospelov,
  Phys.\ Rev.\ Lett.\  {\bf 98}, 231301 (2007)
  [arXiv:hep-ph/0605215];
  M.~Pospelov, J.~Pradler and F.~D.~Steffen,
  arXiv:0807.4287 [hep-ph].

\bibitem{FY}
  M.~Fujii and T.~Yanagida,
  Phys.\ Lett.\  B {\bf 549} (2002) 273
  [arXiv:hep-ph/0208191].



\bibitem{Fox:2002bu}
  P.~J.~Fox, A.~E.~Nelson and N.~Weiner,
  JHEP {\bf 0208}, 035 (2002)
  [arXiv:hep-ph/0206096];
  A.~E.~Nelson, N.~Rius, V.~Sanz and M.~Unsal,
  JHEP {\bf 0208}, 039 (2002)
  [arXiv:hep-ph/0206102];
  G.~D.~Kribs, E.~Poppitz and N.~Weiner,
  arXiv:0712.2039 [hep-ph];
G.~D.~Kribs, A.~Martin and T.~S.~Roy,
  arXiv:0807.4936 [hep-ph].

\end{thebibliography}
\end{document}